\documentclass[aps,pra]{revtex4}
\usepackage{epsfig}
\newcommand{\be}{\begin{equation}}
\newcommand{\bea}{\begin{eqnarray}}
\newcommand{\ee}{\end{equation}}
\newcommand{\eea}{\end{eqnarray}}
\newcommand{\eq}[1]{Eq.~\ref{#1}}

\begin{document}
\title{Long range Casimir force induced by transverse electromagnetic modes}
\author{Ezequiel \'Alvarez and Francisco D.\ Mazzitelli}

\affiliation{ Departamento de F\'\i sica {\it J.J. Giambiagi}, FCEyN UBA,
Ciudad Universitaria, Pabell\' on I, 1428 Buenos Aires, Argentina}

\begin{abstract}
We consider the interaction of two perfectly conducting plates of arbitrary shape
that are inside a non-simply connected cylinder with transverse section of the same shape. We show that
the existence of transverse electromagnetic  (TEM) modes produces a Casimir force that decays
only as $1/a^2$, where $a$ is the distance between plates. The TEM force does not depend on the 
area of the plates and dominates at large distances over the force produced
by the   transverse electric (TE) and 
transverse magnetic (TM) modes, providing in this way a physical realization of the $1+1$ dimensional Casimir effect. For the particular case of a coaxial circular cylindrical cavity, we compute the TE, TM and TEM contributions to the 
force, and find the critical distance for which the TEM modes dominate.

\end{abstract}


\maketitle

\section{Introduction}

In the last years, there has been an increasing interest in the Casimir effect \cite{reviews}. The new 
generation of experiments \cite{experiments} allowed a precise determination of the Casimir 
force, and stimulated theoretical calculations of the forces
for different geometries, including finite temperature and conductivity
corrections. 

The analysis of the dependence of the Casimir force with the geometry is 
therefore both  of theoretical and experimental relevance. In this paper, we will point out that, in non-simply connected electromagnetic cavities, the presence of TEM modes produces an additional contribution to the Casimir force, that it is independent of the section of the cavity and decays slower than the contributions of TE and TM modes. As far as we know, this is the first example in the literature that illustrates the relevance of the TEM modes 
in the {\it static} Casimir effect (for its relevance in the {\it dynamical} Casimir effect see \cite{TEMdce1,TEMdce2} ).

Concretely, we will consider the interaction between two identical perfectly conducting plates that are inside a very long
cylinder of the same section, that is also perfectly conducting. These geometries are usually referred
to as "Casimir pistons" \cite{cavalcanti}, and have received considerable attention recently
\cite{pistons,marachevsky,limteo}. One of the reasons that 
triggered these investigations was the 
reconsideration of the Casimir energy for rectangular boxes, since  repulsive forces
have been predicted when  considering only the zero-point energy of the internal modes of the 
rectangular cavities \cite{paralellep}. The validity of these results has been disputed
\cite{cavalcanti}
for  at least  two related reasons: the omission of the contribution of the external modes, and 
the ambiguity in the renormalization of the divergent quantities (however, there is no consensus in the literature on these issues, see
for instance \cite{geyer}). 
In any case, the advantage of the pistons is that, as long as the surfaces are perfectly conducting,  one can compute the Casimir energy and forces unambiguously and
without considering the external modes  to the cavity.
The new aspect that we will introduce in this paper is the consideration of non-simply connected cavities (see Fig.\ref{display}), allowing for the existence of TEM modes.

The paper is organized as follows. In Section II we will describe the different contributions 
to the zero-point energy in a non-simply connected cavity . We will see that the Casimir energy for TEM modes is 
equivalent to that of a massless scalar field living in  $1+1$ dimensions and  
satisfying Dirichlet boundary conditions on the plates. Moreover, the Casimir energy
for TE and TM modes is equivalent to that of a set of massive, $1+1$-dimensional scalar fields, with the masses determined
by the eigenfrequencies associated to the transverse  section of the cavity. We will discuss the 
behaviour of the force when the distance  between the plates is much larger or much smaller than
the transverse dimensions of the cavity, and conclude that the TEM force dominates above a 
critical distance. In Section III we will present a detailed analysis of the  particular case of a coaxial 
cavity of circular section. We will evaluate the contributions of TE and TM modes to the Casimir energy for this geometry using  a combination of the analytical result for a $1+1$ massive field and Cauchy's theorem
to perform the summation over the effective masses. We will also  compare these contributions with  
that coming from TEM modes,
and find the critical distance as a function of the radii of the inner and outer  cylindrical
shells. Section IV contains our 
final remarks.  
\section{Casimir energy in non-simply connected cavities}

\begin{figure}[h]
\includegraphics[width=.7\textwidth]{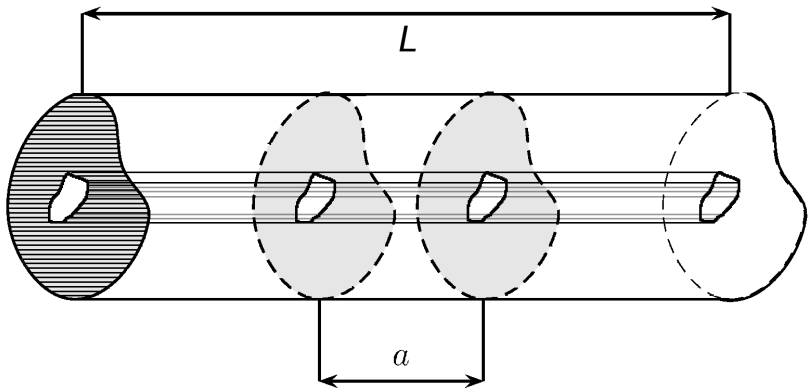}
\caption{Two pistons separated by a distance $a$ inside a non-simply connected cavity (the annular region between two cylinders
of arbitrary section). The system is enclosed between another two plates separated by a distance $L\gg a$.  All surfaces are perfect conductors.  }
\label{display}
\end{figure}

Let us consider a very long electromagnetic cylindrical cavity, with an arbitrary section. We will
assume that the cavity is non-simply connected, i.e. that there is a second cylinder, also
of arbitrary section, inside the larger one (see Fig.\ref{display}). The cavity
is the annular region between the two  cylinders and contains two plates
(pistons) separated by a distance $a$ (the pistons cover only the annular region
between the cylinders). All surfaces are perfectly conducting.
The ${\bf z}$-direction is the axis of the cavity, and we will denote by ${\bf x}_\perp$
the coordinates in the transverse sections.   

At the classical level, the electromagnetic field admits a description in terms of independent TE, TM and TEM modes, which are defined with respect to ${\bf z}$-direction.  This is
possible due to the particular geometries we are considering, that have an invariant section
along the ${\bf  z}$ axis.
The TE and TM electromagnetic degrees
of freedom can be written in terms of two different vector potentials 
${\bf A}_{\rm TE}$ and ${\bf A}_{\rm TM}$ with null divergence
and ${\bf z}$-component \cite{aclar}. The TE electric and magnetic fields are
given by
\begin{eqnarray}
 {\bf E}_{\rm TE}=-\dot{\bf A}_{\rm TE} &;& {\bf B}_{\rm TE}=\nabla\times{\bf A}_{\rm TE},
\label{dualE}
\end{eqnarray}
while the TM fields are given by the dual relations
\begin{eqnarray}
{\bf B}_{\rm TM}=\dot{\bf A}_{\rm TM} &;& {\bf E}_{\rm TM}=\nabla\times{\bf A}_{\rm TM}
\label{dualM} .
\end{eqnarray}
The vector potentials can be written in terms of the so called (scalar) Hertz potentials \cite{TEMdce1,hertzpot}
as
\begin{eqnarray}
{\bf A}_{\rm TE}&=&{\bf\hat z}\times\nabla\phi_{\rm TE} , \nonumber\\
{\bf A}_{\rm TM}&=& {\bf\hat
z}\times\nabla\phi_{\rm TM} ,\label{relations}
\end{eqnarray}
The Hertz potentials $\phi_{\rm TE}$ and $\phi_{\rm TM}$ satisfy the wave equation with Neumann and Dirichlet boundary conditions on the lateral surfaces, respectively, 
and the opposite boundary conditions on the pistons. The eigenfunctions
can be chosen of the form
\begin{equation}
\phi_{\rm TE}(t,z,{\bf x}_\perp )=e^{-iwt}\sin(\frac{n\pi z}{a})\varphi_{\rm TE}({\bf x}_\perp)\, 
\end{equation}
and 
\begin{equation}
\phi_{\rm TM}(t,z,{\bf x}_\perp )=e^{-iwt}\cos(\frac{n\pi z}{a})\varphi_{\rm TM}({\bf x}_\perp)\, ,
\end{equation}
where $n$ is a non-negative integer and  $\varphi_{\rm TE,TM}$ are eigenfunctions of the
transverse Laplacian 
\begin{equation}
\nabla^2_\perp \varphi_{\rm TE,TM} = - \lambda^2 \varphi_{\rm TE,TM}\, .
\label{tlap}
\end{equation}
Therefore, the eigenfrequencies associated to the TE and TM modes are
\begin{equation}
\begin{array}{rcl}
w_{k,n}^{\rm TE}&=&\sqrt{(\frac{n\pi}{a})^2+\lambda_{kN}^2}\nonumber\\
w_{k,n}^{\rm TM}&=&\sqrt{(\frac{n\pi}{a})^2+\lambda_{kD}^2}\,\, ,
\end{array}
\label{freq}
\end{equation}
where   $\lambda_{kN}^2$ and $\lambda_{kD}^2$ are the eigenvalues of Eq.\ref{tlap} when the eigenfunctions satisfy Neumann and Dirichlet boundary conditions,
respectively.

When the cylindrical cavity is non-simply connected, in addition to the TE and
TM modes one should also consider the TEM modes, for which both
the electric and magnetic fields have vanishing $z$ component. Working with the usual vector potential
${\bf A}$, the TEM solutions are of the form
\begin{eqnarray}
{\bf A}({\bf x}_{\perp},z,t)&=&{\bf A}_{\perp}({\bf x}_{\perp})\phi_{\rm TEM}(z,t) ,\nonumber \\
{\bf E}&=& - (\partial_t \phi_{\rm TEM}) \; {\bf A}_{\perp} ,          \nonumber \\
{\bf B}&=& (\partial_z \phi_{\rm TEM}) \; {\bf \hat z} \times {\bf A}_{\perp} .
\label{aperp}
\end{eqnarray}
where $\phi_{\rm TEM}(z,t)$ is an additional scalar field.
The transverse vector potential has vanishing rotor and
divergence, and zero tangential component on the transverse surfaces. Therefore, ${\bf A}_{\perp}$ is a solution of an {\it electrostatic} problem in the two transverse dimensions (in hollow cylindrical cavities the transverse potential vanishes and
TEM  modes do not exist). The scalar field $\phi_{\rm TEM}$ satisfies Dirichlet boundary conditions on the longitudinal boundaries
$z=0$ and $z=a$, and the longitudinal wave
equation $(\partial_t^2-\partial_z^2)\phi_{\rm TEM} = 0$. Thus, the eigenfrequencies of the TEM modes are 
\begin{equation}
w_n^{\rm TEM}= n \pi / a\, .
\label{freqtem}
\end{equation}

In order to obtain the Casimir energy we introduce the regularized quantities
\begin{equation}
E_{\rm reg}(a)=\frac{1}{2}\sum_n w_n^{TEM}e^{-\sigma w_n^{TEM}}+\frac{1}{2}\sum_{n,k} (w_{k,n}^{TE}
e^{-\sigma w_{k,n}^{TE}}+w_{k,n}^{TM} e^{-\sigma w_{k,n}^{TM}} )\equiv
E^{TEM}_{\rm reg}(a)+E^{TE}_{\rm reg}(a)+E^{TM}_{\rm reg}(a)\, ,
\label{naive}
\end{equation}
and 
two additional pistons separated at a very large distance $L$, enclosing the system. The physical Casimir energy is
defined as the difference
\begin{equation}
E(a)=   E_{\rm reg}(a) + 2 E_{\rm reg}(\frac{L-a}{2})-3 E_{\rm reg}(\frac{L}{3})
\label{finite}
\end{equation}
in the limit when the cutoff $\sigma$ tends to zero. Note that one can compute independently the
TE, TM and TEM contributions to the energy. Note also that, as the pistons only
cover the annular region between the cylinders, the internal modes of the smaller cylinder
will be irrelevant for the interaction between plates.

To proceed, we note that the Casimir energy for this geometry is formally equivalent to that of a set of scalar fields living in $1+1$ dimensions and satisfying Dirichlet boundary conditions 
at $z=0$ and $z=a$. Indeed, Eq.\ref{freqtem} implies that the TEM Casimir energy is equivalent  to that of a massless
scalar field, and the result is very well known \cite{emassless}
\begin{equation}
E^{TEM}(a)=-\frac{\pi}{12 a}\, .
\label{Etem}
\end{equation}
From Eq.\ref{freq} we see that the TE and TM Casimir energies correspond to that of a set of massive scalar fields, with masses given by 
$\lambda_{kN}$ and $\lambda_{kD}$. The Casimir energy $E_m$ for a field of mass $m$ in $1+1$ dimensions
has been computed previously by many authors \cite{Emassive} using different methods of regularization (see in particular Ref. \cite{limteo} for a calculation with
an exponential cutoff). It reads \cite{pervsdir}
\begin{equation}
E_m(a)=-\frac{1}{2\pi} \sum_{l=1}^{+\infty}   
\frac{m K_1(2l m a)}{l} \, ,
\label{emassive1}
\end{equation}
where $K_1$ is the modified Bessel function of the second kind.
Using this result and the analogy between the TE and TM eigenfrequencies (\eq{freq}) with the eigenfrequencies of massive scalar fields in $1+1$ dimensions, we can easily obtain the TE and TM contributions to the Casimir energy in the cylindrical cavity
\be
E^{TE}(a)+E^{TM}(a)=-\frac{1}{2\pi} \sum_{l=1}^{+\infty} 
\left(   
\sum_{\lambda_{kN}} \frac{\lambda_{kN} K_1(2l\lambda_{kN} a)}{l} +
\sum_{\lambda_{kD}} \frac{\lambda_{kD} K_1(2l\lambda_{kD} a)}{l} 
\right)\, .
\label{mara}
\ee
This equation has been previously obtained in Ref. \cite{marachevsky} using a different method.
We stress that the formula is valid for a cavity of arbitrary section. The energy can in principle be computed through a numerical evaluation of the eigenvalues 
of the transverse Laplacian. Alternatively, as we will describe in the next section, the summation
over the eigenvalues can be performed using Cauchy's theorem. The force between pistons is
easily obtained taking the derivative of the energy with respect to $a$.

Let us now discuss some generic properties of the different 
contributions to the Casimir energy. At small distances, when the separation between pistons is much smaller that the transverse dimensions of the cavity, one expects the proximity force approximation (PFA) to describe accurately the contributions of TE and TM modes. Indeed, 
using heat kernel techniques it can be shown \cite{marachevsky} that, in this limit
\begin{equation}
E^{TE}(a)+E^{TM}(a)\approx-\frac{\pi^2}{720}\frac{A}{a^3}\, ,
\label{pfa}
\end{equation}
where $A$ is the area of the transverse sections. This is of course the well known result for parallel plates. 
It is worth to stress that the geometric properties of the transverse section, as the area, are contained in the eigenvalues $\lambda_{kD}$ and $\lambda_{kN}$, which play the role of the masses of the fields in the $1+1$ dimensional analogy.

On the other hand, in the opposite limit we have $\lambda a\gg 1$. The TE and TM  contributions to the Casimir energy are 
dominated by the lowest eigenvalue $\lambda_{MIN}$, and have the typical exponential
suppression associated to massive fields, i.e.
\begin{equation}
E^{TE}(a)+E^{TM}(a)\approx -\kappa\sqrt{\frac{\lambda_{MIN}}{16\pi a}}e^{-2\lambda_{MIN}a}\, ,
\label{exp}
\end{equation}
where $\kappa$ is the multiplicity of the eigenvalue.

From these results, we conclude that, at small distances, the total Casimir energy is dominated by the TE and TM contributions: it behaves as $1/a^3$ as for parallel plates, and it is proportional to the area of the pistons. For distances larger than a critical value,  $a>a_c$, the TEM is the leading contribution, and gives a long range Casimir energy that decays only as $1/a$. This contribution, typical of a massless scalar field in $1+1$ dimensions, is non-extensive, i.e.
does not depend on the area of the pistons. The value of $a_c$ depends of course of the particular form of the transverse
section. 

There are some additional properties which can be obtained using dimensional analysis. 
Let us denote by $l_1$ and $l_2$ the typical lengths associated to the sections of the  internal and  
external cylinders of the cavity, respectively.
On dimensional grounds we expect 
\begin{equation}
E^{TE}+E^{TM}=\frac{1}{a}f\left(\frac{a}{l_2},\frac{l_1}{l_2}\right)\, .
\end{equation}
If the critical distance is defined by 
\begin{equation}
 f\left(\frac{a_c}{l_2},\frac{l_1}{l_2}\right)=-\frac{\pi}{12}
\end{equation}
then we have that
\be
 a_c=l_2\, g(l_1/l_2).
\label{ac}
\ee
In the particular  case $l_1\ll a_c,\ l_2$, on physical grounds we expect the functions $f$ and $g$ defined above to have well defined limits: 
\be
E^{TE}+E^{TM}\approx \frac{1}{a}f\left(\frac{a}{l_2},0\right)\,,\,\, 
 a_c=l_2\, g(0).
\label{ac limit}
\ee
Indeed, this  limit can be achieved by inserting a thin wire inside a hollow cavity, so that the TE and TM modes
of the hollow cavity are not disturbed, and so the TE and TM contributions to the zero-point energy
are almost independent of the presence of the wire. In this situation
the critical distance becomes a linear function 
of $l_2$. We will confirm this property in the particular example described in the next section.

The existence of the long range Casimir TEM force is of conceptual interest. 
One can wonder whether is it also relevant from an experimental perspective,
i.e. if there is a chance of measuring this force in future experiments.
There are two major limitations: on the one hand, being non-extensive in the area
of the plates, the absolute value of the force is very small, and therefore it could only be measured
at extremely short distances. Moreover, in this regime, the Casimir force would be dominated
by the TE and TM contributions, unless the area of the plates is also small.  
We therefore address the question of 
whether the TEM contribution to the force can be a significant fraction ($\alpha$)
of the total
Casimir force in the PFA regime, i.e. we are interested in a configuration in which  $F^{TEM}\geq 
\alpha(F^{TE}+F^{TM})$.
For the sake  of concreteness we study the case where the geometry of the plates is such that its area may be written as $b(l_2^2-l_1^2)$,  where $b$ is some coefficient (this occurs, for instance, with a circle and any regular polygon).  Within these assumptions, the conditions imposed to the forces become the following conditions in the parameters,
\be
b \ll \frac{A}{a^2} \leq \frac{20}{\pi\alpha} \, ,
\label{condition}
\ee
where the first inequality follows from the validity of the PFA.
This shows that, in principle,  there is a region in the parameter space where the TEM force could be 
a significant fraction of the force, even in the proximity limit. However, we stress again that the smallness
of the TEM force implies that its measurement is presently extremely difficult.
\section{Coaxial cylindrical cavity of circular section}

In this Section we present and solve an explicit example of a non-simply connected cavity.  We compute the Casimir energy 
for a cavity formed by two concentric perfectly conducting circular cylinders closed by two (also non-simply connected) plates in its extremes.  The configuration is similar to that in Fig.\ref{display} but with both inner and outer cylinders having circular sections of radii $r_1$ and $r_2$, respectively.  We also study the Casimir force between the plates.

As discussed in the previous Section, the vacuum energy of this non-simply connected cavity will have contributions coming from the TEM modes, besides the usual TE and TM modes.  The existence of TEM modes can be 
confirmed by obtaining explicitly the transverse vector potential ${\bf A}_\perp$ defined in Eq.\ref{aperp}, that for this
particular geometry reads
\begin{equation}
{\bf A}_{\perp} =\frac{ {\hat{\bf x}_\perp }}{ \vert{\bf x}_\perp\vert }\,\, .
\end{equation}
The TEM modes contribution to the energy is 
given by Eq.\ref{Etem}, and is independent of $r_1$ and $r_2$, as stated above.

On the other hand, the TE and TM modes contribution do not have such a simple expression and do depend on $r_1$ and $r_2$.  To obtain this contribution, we may start from the already finite expression given in \eq{mara} and use Cauchy's theorem to convert the sum over $\lambda_{D}$'s and $\lambda_{N}$'s into a closed path integral in the complex plane of an appropriate function. 

If $f(z)$ is a function with ''1''-valued simple poles at $z=\lambda_{kD}$ and $z=\lambda_{kN}$ for all $\lambda_{kD,kN}$ then we may compute the sum in $\lambda=\lambda_{kD,kN}$ in \eq{mara} as a Cauchy integral,
\be
\sum_{\lambda} \frac{\lambda K_1(2 l \lambda a)}{l} = 2\pi i \int_{{\cal C}} z \frac{ K_1(2 l z a)}{l} f(z),
\ee 
if the contour ${\cal C}$ encloses all the poles of $f(z)$ in $z=\lambda_{kD,kN}$ and the function $K_1(2lza)$ is analytic in the interior of the curve.  The explicit form of $f(z)$ comes out after observing that in the case of the cavity between two perfectly conducting concentric cylinders of radius $r_1$ and $r_2$, the Dirichlet and Neumann eigenfrequencies --which correspond to the TM and TE modes, respectively-- are all the solutions of (see \cite{javier}) 
\be
\begin{array}{rcl}
J_n(\lambda r_1) N_n (\lambda r_2) - J_n(\lambda r_2) N_n(\lambda r_1) &=& 0, \\
J'_n(\lambda r_1) N'_n (\lambda r_2) - J'_n(\lambda r_2) N'_n(\lambda r_1) &=& 0,
\end{array}
\label{cc}
\ee
with $n$ any integer number.  (Notice that if $\lambda$ is a solution then $-\lambda$ is also a solution, but since both correspond to the same eigenfunction we may keep only the $\lambda>0$ solutions to avoid double counting.)  From here it is easy to see that
\be
f(z)=\sum_{n} \frac{d}{dz} \ln 
\left[
\left(
J_n(z r_1) N_n (z r_2) - J_n(z r_2) N_n(z r_1)
\right)
\left(
J'_n(z r_1) N'_n (z r_2) - J'_n(z r_2) N'_n(z r_1)
\right)
\right]
\label{suma}
\ee
satisfies the above requirements. 

In order to choose the contour ${\cal C}$ we observe that $K_1(z)$ is singular at $z=0$ but is analytic for $Re(z)>0$, where goes to zero as $z^{-1/2}\,e^{-Re(z)}$ for large $Re(z)$.  Moreover, since the contour must enclose the real positive axis beginning in $\lambda_{MIN}$ (the minimum of the solutions of \eq{cc} for all $n$) we choose the contour to be a {\it pizza slice} with its vertex at $z=\lambda_{MIN}/2$, angle $0<\phi<\pi/2$, and centered in the real axis:
\bea
{\cal C} = \lim_{L\to\infty}
\left\{
\begin{array}{lll}
z=\frac{\lambda_{MIN}}{2} +\rho\, e^{-i\phi/2} & \rho \in (0,L)& \\ 
z= \frac{\lambda_{MIN}}{2} + L \, e^{i\theta} & \theta \in (-\phi/2,+\phi/2)&\\
z= \frac{\lambda_{MIN}}{2} + \rho \, e^{+i\phi/2} & \rho \in (L,0)& \\
\end{array}
\right.
\eea

Once $f(z)$ and ${\cal C}$ have been correctly chosen, we have an explicit expression for the TE+TM Casimir energy in terms of a double sum and a closed path integral,
\be
E^{TE}+E^{TM}= - i \sum_{l=1,n=-\infty}^{\infty} \int_{{\cal C}} z \frac{ K_1(2 l z a)}{l} f_n(z),
\label{tetm}
\ee
where $f_n(z)$ is each term in the sum in $f(z)$ (see \eq{suma}).

To compute explicitly \eq{tetm} we need to truncate the $l$ and $n$ sum according to a given precision, and compute numerically the integral in the upper and lower segments of ${\cal C}$  --which are essentially the same--, since the contribution in the arc of radius $L$ goes to zero.  The criteria used to truncate the sum is best analyzed in \eq{mara}, where the sum is exponentially damped by the Bessel function when its argument grows.  In fact, we may divide in \eq{mara} the sum in $\lambda$ as different sums for each $n$ (see \eq{cc}), then we define $\lambda_{min}(n)$ as the minimum of the $\lambda$'s for a given $n$, and then for each $n$ we keep $l$'s such that $2l \lambda_{min}(n) a < D$.  Here $D$ is chosen such that all the thrown away terms in the sum are damped by at least an $e^{-D}$ factor.  On the other hand, to truncate the sum in $n$, we set $l=1$ and we choose $n$ such that $2 \lambda_{min}(n) a<D$.  This criteria should give a precision of order $e^{-D}$ to the final result in the sum. In our calculations we have taken $D=8$ which is enough for our purposes.  

It is worth noticing at this point that if we would have performed directly the sum in \eq{mara} over all relevant $\lambda$'s instead of using the Cauchy integral approach, then we would have had to study the roots of \eq{cc} for each $n$ and keep only those which satisfy $2 l \lambda a<D$ for each $l$.  Although more difficult, this would have also been a possible approach.

The TE+TM Casimir energy has been numerically computed using \eq{tetm} for different cylinders radii ($r_1$ and $r_2$) and distance between the plates ($a$).  We have also computed the Casimir force deriving \eq{tetm} with respect
to $a$ and computing numerically the resulting expression: 
\begin{equation}
 F^{TE}+F^{TM}=  -i \sum_{l=1,n=-\infty}^{\infty} \int_{{\cal C}} z^2 \left( K_0(2 l z a)+K_2(2 l z a)\right) f_n(z).
\label{ftetm}
\end{equation}

\begin{figure}[h]
\includegraphics[width=.7\textwidth]{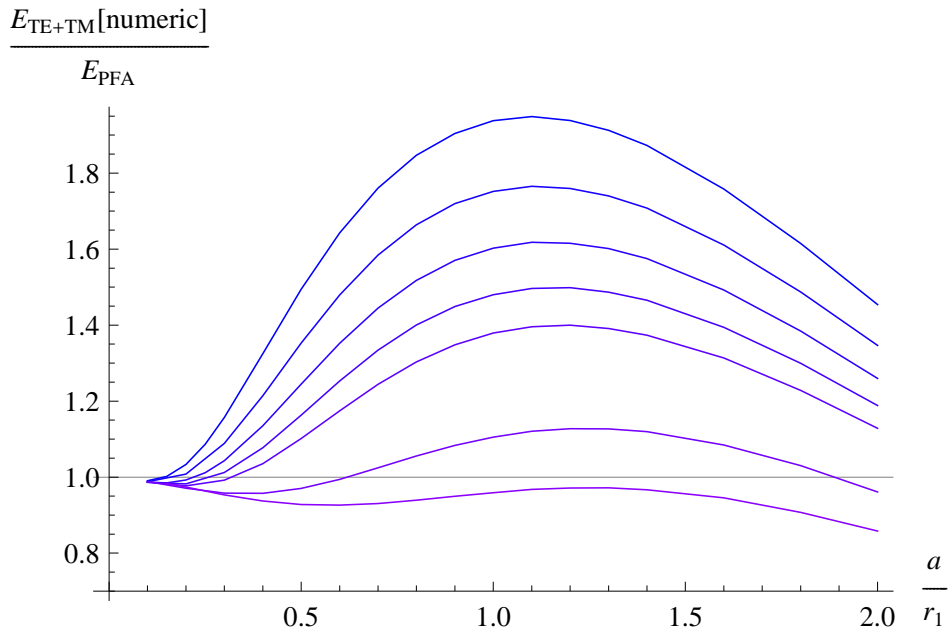}
\caption{Ratio of the numerically computed TE+TM Casimir energy to the expected proximity behaviour (\eq{pfa}) in the small-$a$ region.  Our calculations reach $a/r_1 \approx 0.1$ which is enough for our purposes.  The lines correspond, from lower to upper, to $r_2/r_1=2,\ 1.8,\ 1.6,\ 1.55,\ 1.5,\ 1.45$ and $1.4$. }
\label{a-small}
\end{figure}

As a check for the numerical TE+TM calculation we have corroborated that its behaviour for small and large $a$ corresponds to the expected proximity (Eq. \ref{pfa}) and exponential (Eq. \ref{exp}) behaviours, respectively.   In both cases we find, as expected, a convergence to unit in the ratio of the numerical
energy and its expected asymptotic behaviour. In Fig.\ref{a-small} we show this convergence in the proximity limit,
which is the more complex from a numerical point of view, since it involves the summation of a large number of modes.

\begin{figure}[h]
\includegraphics[width=.7\textwidth]{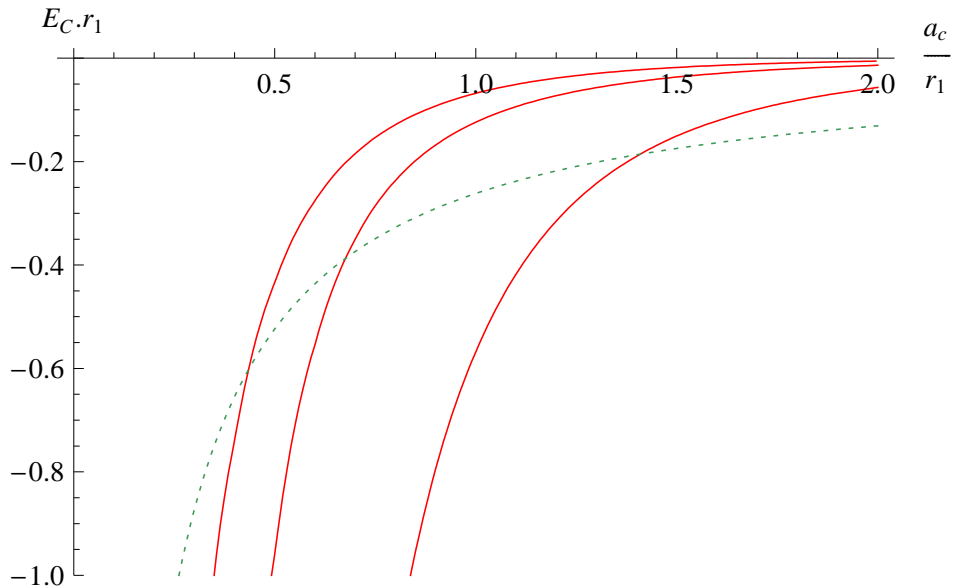}
\caption{TEM (dotted) and TE+TM (solid) contributions to the Casimir energy for different $r_2/r_1$ configurations as a function of the distance $a$ (in units of $r_1$) between the plates.  From left to right the solid lines correspond to $r_2/r_1=1.2,\ 2$ and $4$, and the corresponding critical distances are $a_c=0.44\, r_1,\ 0.68\,r_1$ and $1.41\,r_1$, respectively.  Observe that the TEM contribution is independent of $r_2/r_1$.}
\label{2plots}
\end{figure}

In order to explore the TE+TM to TEM transition, we have studied the ranges $r_2=1.1\, r_1$ to $r_2=40 \,r_1$ and focused on the region of $a$'s where the TE+TM energy (or force) equals the TEM energy (or force).  As usual, the numerical calculation becomes harder when the surfaces become closer, i.e. when $r_2$ approaches $r_1$.
In Figure \ref{2plots} we have plotted the TE+TM Casimir energy as a function of $a$ together with the TEM Casimir energy for three different $r_2/r_1$ configurations.  As expected, the TE+TM energy dominates in the small-$a$ region, but the TEM energy dominates when the distance between the plates is larger than the critical distance $a_{c}$. (Note 
that given $r_1$ and $r_2$ there are two different critical distances, one for the energy and another
for the force. With no risk of confusion we use the notation $a_c$ for both of them.)

\begin{figure}[h]
\includegraphics[width=.7\textwidth]{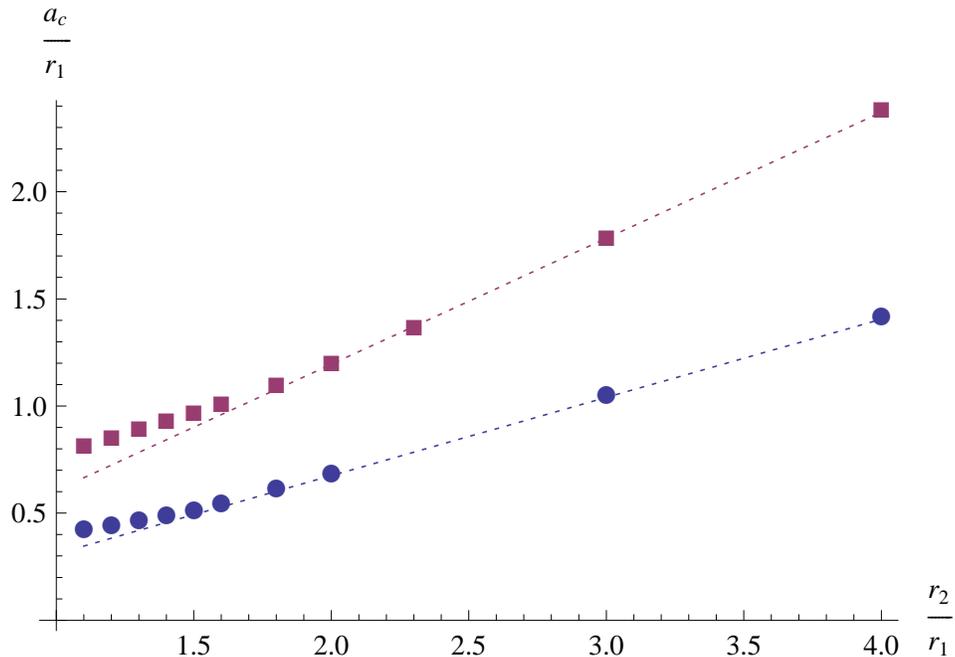}
\caption{Critical distance (in $r_1$ units) where the Casimir energy (lower plot) or force (upper) of the TE+TM modes equals that of the TEM modes for different $r_2/r_1$ configurations.  As it can be seen, the dependence of $a_{c}$ with $r_2$ becomes rapidly linear. The slopes of the asymptotes are $0.36$ and $0.61$. }
\label{aeq}
\end{figure}

We have plotted in Figure \ref{aeq} the Casimir energy and force critical distance $a_{c}$ for several radii ratios $r_2/r_1$.  As it can be seen in the plot, the dependence of $a_{c}$ with $r_2$ becomes rapidly linear. According to
Eq.\ref{ac}, this means that $a_c= r_2\, g(r_1/r_2)$ approaches its limiting value $a_c\approx r_2\, g(0)$ for
$r_2>2r_1$. In this case we find $a_c\approx 0.36 \, r_2$,  for the critical distance associated to the energy.

\begin{figure}[h]
\includegraphics[width=.7\textwidth]{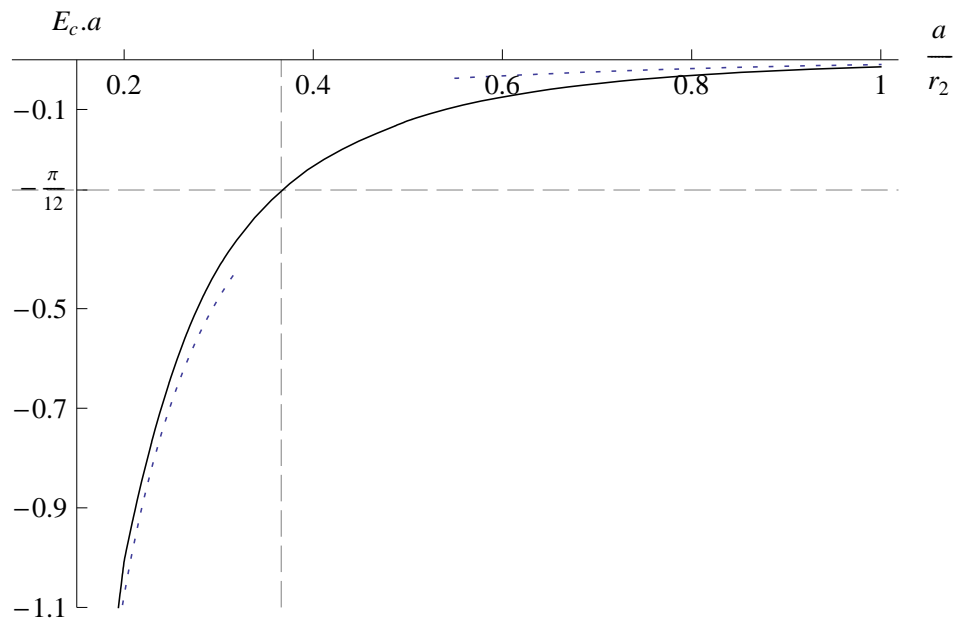}
\caption{Casimir energy for two plates separated a distance $a$ in a {\it hollow} circular cylinder of radius $r_2$.  The expected PFA and exponential behaviour for small and large $a/r_2$, respectively, is plotted in dotted lines. The dashed lines show that, as expected, this energy equals $E^{TEM}=-\pi/12 a$ at $a/r_2 \approx\, 0.36$ (see text and Fig. \ref{aeq}).}
\label{hollow}
\end{figure}

As mentioned at the end of Section II, the linear relation between $a_c$ and $r_2$   corresponds to 
the physical situation in which the 
TE+TM Casimir energy of the non-simply connected  cavity approaches that of a simply connected
one of radius $r_2$. Therefore, one should be able to obtain the coefficient $g(0)$ by a comparison of 
the TEM Casimir energy  with  that
of a hollow cylindrical
cavity of radius $r_2$,  which is given by Eq.\ref{tetm} with
\be
f_n(z)= \frac{d}{dz} \ln 
\left(
J_n(z r_2) J'_n (z r_2)\right)
\ee
We have checked that this is indeed the case. The Casimir energy for the hollow cavity, obtained again using Cauchy's theorem, is plotted in Fig \ref{hollow}.  As expected, this energy interpolates between the proximity result at short distances, and the exponential
behaviour at long distances. The number $g(0)$ is determined by the value of $a/r_2$ for
which the energy of the hollow cavity equals $-\pi/12 a$. In this way we obtain
 $g(0)\approx 0.36$, that coincides with the slope of the linear relation between the critical 
 distance for the energy and $r_2$ presented in Fig. \ref{aeq}.
 
\section{Conclusions}

In this paper we have described a geometry in which the existence of  TEM modes
induce a long range Casimir interaction. In particular, 
we have shown that the electromagnetic Casimir force between two parallel plates inside a 
non-simply connected cylinder is essentially
given by
the sum of the Casimir forces for $1+1$ scalar fields with different masses. For TE and TM modes,
the masses are given by the eigenvalues of the 
Laplacian on a $z=const$ section of the cylinder, with the appropriate boundary conditions,
and  
are non-vanishing. The opposite happens to TEM modes, whose zero-point energy 
corresponds to  a massless field, and this is the reason of the different
qualitative behaviour of their contribution  to the force.  On the one hand, the TEM force scales as 
$1/a^2$
at all distances and does not depend on the area of the plates. On the other hand, in the short distance limit TE and TM forces reproduce
the parallel plates result proportional to $A/a^4$, and in the long distance limit they are 
exponentially suppressed due to the finite size of the plates (or, in the equivalent picture,
to the non-vanishing effective masses). As a consequence, TE and TM modes dominate at 
short distances, while  TEM modes do it at long distances. The critical distance 
where both contributions are balanced depends of
course on the form of the section of the cylinder, and decreases with its area, as we
explicitly showed in the particular example of a coaxial circular cylindrical cavity. The summation
over the effective masses to compute the TE and TM contributions of the force 
for this particular case was performed
using Cauchy's theorem, starting from the renormalized Casimir energy for a single massive
field in $1+1$ dimensions.

Throughout the paper we considered the Casimir energy for perfect conductors at zero temperature.
It would be interesting to generalize these results to take into account the
combined  effects of finite 
conductivity at nonzero temperature.
In this context, it is worth  remarking that the dominance of the TEM modes
over the TE and TM contributions to the Casimir force is also valid at a  nonzero temperature $T$,
at least  for perfect 
conductors. In fact, it has been shown \cite{marachevsky} that at sufficiently long distances ($\lambda_{MIN} a\gg 1$
and  $aT\gg 1$), the Casimir force for TE and TM modes is proportional to $\lambda_{MIN} T e^{-2\lambda_{MIN} a}$,
i.e. is exponentially suppressed. On the other hand, in the same situation the TEM force is
proportional to $T/a$ \cite{temp}. 
 
While the existence of the long-range TEM Casimir force is of conceptual interest,
it would be very difficult to measure it. Indeed, as discussed at the end of Section II, 
this contribution to the force
does not depend on the area of the plates, its absolute value is extremely small, and therefore it would be 
measurable only at very short distances
with the present technology. However, in this regime the TE and
TM forces would  be much larger than TEM force, unless the area of the pistons is sufficiently 
small.  Therefore, the measurement of the TEM Casimir force should involve thin rings at
short distances. A rather difficult experiment, indeed. 

\section{acknowledgments}
This work was supported by Universidad de Buenos Aires, CONICET and ANPCyT. We would like to thank
F. Fiorini for asking the right question.


\begin{thebibliography}{99}
\bibitem{reviews} For recent reviews, see 
  M.~Bordag, U.~Mohideen and V.~M.~Mostepanenko,
  Phys.\ Rept.\  {\bf 353}, 1 (2001)
  [arXiv:quant-ph/0106045];
; K.A. Milton, J. Phys. A {\bf 24}, R209 (2004); S.K. Lamoreaux, Rep.
 Prog. Phys. {\bf 68}, 201 (2005).
 \bibitem{experiments}
 S.K. Lamoreaux, Phys. Rev. Lett. {\bf 78}, 5 (1997);
U. Mohideen and A. Roy, Phys. Rev. Lett. {\bf 81}, 4549 (1998); G. Bressi, G. Carugno, R. Onofrio, and G. Ruoso, Phys. Rev. Lett.
{\bf 88}, 041804 (2002); B.W. Harris, F. Chen, and U. Mohideen, Phys. Rev. A {\bf 62}, 052109 (2000);
T. Ederth, Phys. Rev. A {\bf 62}, 062104 (2000);
H.B. Chan, V.A. Aksyuk, R.N. Kleiman, D.J.Bishop, and F. Capasso, Science 291, 1941 (2001); H. B. Chan, V.A. Aksyuk, R.N. Kleiman, D.J. Bishop, and F. Capasso, Phys. Rev. Lett. {\bf 87}, 211801 (2001);
D. Iannuzzi, I. Gelfand, M.Lisanti, and F. Capasso, Proc. Nat. Ac. Sci. USA 101, 4019 (2004); R.S. Decca, D. Lopez, E. Fischbach, and D.E. Krause, Phys. Rev. Lett. {\bf 91}, 050402 (2003);
R.S. Decca et al., Phys. Rev. Lett. {\bf 94}, 240401 (2005);
R.S. Decca et al., Annals of Physics {\bf 318}, 37 (2005); 
R.S. Decca et al., Eur. Phys. J. C {\bf 51}, 963, (2007);
  R.~S.~Decca, D.~Lopez, E.~Fischbach, G.~L.~Klimchitskaya, D.~E.~Krause and V.~M.~Mostepanenko,
  Phys.\ Rev.\  D {\bf 75}, 077101 (2007)
  [arXiv:hep-ph/0703290];
F.~Chen, G.L.~Klimchitskaya, V.M.~Mostepanenko, and U.~Mohideen,
  Phys.\ Rev.\ B {\bf 76}, 035338 (2007).
\bibitem{TEMdce1} M. Crocce, D.A. Dalvit, F.C. Lombardo and F.D. Mazzitelli, 
J. \ Opt.\ B: Quantum Semiclass. Opt. {\bf 7}, S32 (2005).
\bibitem{TEMdce2} P.A. Maia Neto, 
 J. \ Opt.\ B: Quantum Semiclass. Opt. {\bf 7}, S86 (2005).
\bibitem{cavalcanti}
  R.~M.~Cavalcanti,
  Phys.\ Rev.\  D {\bf 69}, 065015 (2004)
  [arXiv:quant-ph/0310184].
\bibitem{pistons}
  M.~P.~Hertzberg, R.~L.~Jaffe, M.~Kardar and A.~Scardicchio,
  Phys.\ Rev.\ Lett.\  {\bf 95}, 250402 (2005)
  [arXiv:quant-ph/0509071];
  G.~Barton,
  Phys.\ Rev.\  D {\bf 73}, 065018 (2006);
S.A. Fulling and J.H. Wilson, Phys. Rev. A {\bf 76}, 012118 (2007);
  A.~Edery,
  Phys.\ Rev.\  D {\bf 75}, 105012 (2007)
  [arXiv:hep-th/0610173].
X. Zhai, X.  Li, Phys.Rev. D {\bf 76}, 047704 (2007);
V.N. Marachevsky, J.Phys.A {\bf 41}, 164007 (2008);
  K.~Kirsten and S.~A.~Fulling,
  arXiv:0901.1902 [hep-th];
  A.~Edery and V.~Marachevsky,
  Phys.\ Rev.\  D {\bf 78}, 025021 (2008)
  [arXiv:0805.4038 [hep-th]];
  A.~Edery and I.~MacDonald,
  JHEP {\bf 0709}, 005 (2007)
  [arXiv:0708.0392 [hep-th]].
  \bibitem{marachevsky}
  V.~N.~Marachevsky,
  Phys.\ Rev.\  D {\bf 75}, 085019 (2007)
  [arXiv:hep-th/0703158].
\bibitem{limteo} S.C. Lim and  L.P.  Teo,  Arxiv preprint arXiv:0807.3613 (2008).
\bibitem{paralellep} W. Lukosz, Physica (Uthrecht) {\bf 56}, 109 (1971);
S. Ambjorn and S. Wolfram, Ann. Phys. (NY) {\bf 147}, 1 (1983); G.J. Maclay, Phys. Rev. A {\bf 61},
052110 (2000).
\bibitem{geyer}
B. Geyer, G.L. Klimchitskaya and V.M. Mostepanenko, Eur. Phys. J . C {\bf 57}, 823
(2008). 
\bibitem{aclar} P.A. Maia Neto, J. Phys. A {\bf 27}, 2164 (1994); D.F. Mundarain and P.A. Maia Neto,
Phys. Rev. A {\bf 57}, 1379 (1998).
\bibitem{hertzpot} A. Nisbet, Proc. of the Royal Soc. of London. Series A, Mathematical
and Physical Sciences, Vol. 231, 250 (1955). See also Ref.\cite{TEMdce1}.
\bibitem{emassless} See for instance, Bordag et al in Ref.\cite{reviews}.
\bibitem{Emassive} See for instance {\it The Casimir effect}, K.A. Milton, World Scientific (2001),
Chapter 2. 
\bibitem{pervsdir} This energy is usually computed for periodic boundary conditions in $1+1$
dimensions. Eq.
\ref{emassive1} contains an additional $1/2$ factor due to the fact that we are considering
Dirichlet or Neumann boundary conditions, and therefore the eigenfrequencies in Eq.\ref{freqtem}
 involve only  positive integers.
\bibitem{javier}
  F.~D.~Mazzitelli, M.~J.~Sanchez, N.~N.~Scoccola and J.~von Stecher,
  Phys.\ Rev.\  A {\bf 67}, 013807 (2003)
  [arXiv:quant-ph/0209097].
\bibitem{temp} This can be checked by evaluating the finite temperature corrections to the Casimir 
force for a 
massless scalar field in $1+1$ dimensions. See for instance 
  S.~C.~Lim and L.~P.~Teo,
  arXiv:0808.0047 [hep-th];
K. Milton,  hep-th/9901011.
\end{thebibliography}
\end{document}